\documentclass[showpacs,floatfix,aps,prb,reprint,superscriptaddress]{revtex4-1}
\usepackage{xfrac}
\usepackage{amssymb}
\usepackage{braket}
\usepackage{graphicx}
\usepackage{etoolbox}
\usepackage{amsfonts} 
\usepackage[fleqn]{amsmath} 
\usepackage[utf8]{inputenc} 
\usepackage{mathtools}
\usepackage{xcolor}
\usepackage{bm} 

\usepackage{tikz}
\usetikzlibrary{arrows, decorations.markings}
\begin{document}

\title{Effective and accurate representation of extended Bloch states on finite Hilbert spaces}
\author{Luis A. Agapito} 
\email{luis.agapito@gmail.com}
\affiliation {Department of Physics, University of North Texas, Denton, TX 76203, USA}
\affiliation {Center for Materials Genomics, Duke University, Durham, NC 27708, USA}
\author{Andrea Ferretti} \affiliation {CNR-NANO S3 Center, Istituto Nanoscienze, I-41125, Modena, Italy}
\author{Arrigo Calzolari} 
\affiliation {Department of Physics, University of North Texas, Denton, TX 76203, USA}
\affiliation {CNR-NANO S3 Center, Istituto Nanoscienze, I-41125, Modena, Italy}
\author{Stefano Curtarolo}
\affiliation {Center for Materials Genomics, Duke University, Durham, NC 27708, USA}
\affiliation {Department of Mechanical Engineering and Materials Science, Duke University, Durham, NC 27708, USA}
\author{Marco \surname{Buongiorno Nardelli}} 
\email{mbn@unt.edu}
\affiliation {Department of Physics, University of North Texas, Denton, TX 76203, USA}
\affiliation {Center for Materials Genomics, Duke University, Durham, NC 27708, USA}
\date{\today}

\begin{abstract}
We present a straightforward, noniterative projection scheme that can represent the electronic ground state of a periodic system on a finite atomic-orbital-like basis, up to a predictable number of electronic states and with controllable accuracy.
By co-filtering the projections of plane-wave Bloch states with high-kinetic-energy components, the richness of the finite space and thus the number of exactly-reproduced bands can be selectively increased at a negligible computational cost, an essential requirement for the design of efficient algorithms for electronic structure simulations of realistic material systems and massive high-throughput investigations.
\end{abstract}


\maketitle
\section{introduction}
The electronic structure of solids is commonly described using plane waves (PW) basis functions, which  represent naturally the Fourier algebra of periodic systems and whose completeness is easily improvable up to any desirable accuracy. However, their delocalized character is often not appropriate for the description of highly localized electronic systems unless a very large number of basis functions is used. For these reasons, the development of minimal-space solutions such as atomic-orbital (AO) Bloch sums, capable of capturing with satisfactory accuracy the properties of solids and molecules on finite Hilbert spaces, has been central to methodological developments in quantum chemistry and solid state physics since the 70's ~\cite{Chadi1977,Sanchez-Portal1995}.

AO representations are desirable not only for computational accuracy and finiteness of the basis set, but also to gain a better chemical interpretation of the quantum-mechanical wavefunction. They are essential in a gamut of applications such as: construction of model Hamiltonians for correlated-electrons and magnetic systems; dynamical mean-filed theory \cite{Lechermann2006}; evaluation of quantum transport properties \cite{[{ \textsc{WanT} code by A. Ferretti, L. Agapito, A. Calzolari, and M. Buongiorno Nardelli, \texttt{http://www.wannier-transport.org}; }]Calzolari2004}; design of semiempirical potentials for solids \cite{Aguado2003} and biomolecules \cite{Sagui2004}; calculation of exact-exchange integrals; and applications within linear scaling of density-functional theory \cite{Ordejon1996}, coupled cluster \cite{Schutz2001}, quantum Monte Carlo \cite{Williamson2001} and the \textsc{gw} \cite{Umari2009} methods. More generally, localized-space representations are increasingly in demand as novel materials with stronger electron localization and correlation are vigorously sought for their rich physical and chemical properties \cite{Curtarolo2013}. Moreover, they allow the calculation of the electronic states of materials on ultra dense $\mathbf{k}$-space grids for accurate Brillouin zone (BZ) integrations, an essential requirement for the high-throughput computational materials applications central to the mission of the Materials Genome Initiative \cite{Curtarolo2013,MGIeditorial}.

In the last decade, formidable efforts towards this goal have resulted in a variety of methodologies using, for instance, Muffin-tin orbitals of arbitrary order (\textit{N}MTO) \cite{Andersen2000} or maximally-localized Wannier functions (MLWF) \cite{Marzari2012} to construct minimal Hilbert spaces. The MLWF method stands as the norm for maximal localization of the real-space basis starting from pseudo-potential plane-wave (PW) calculations; however, it is not straightforward either to decide the appropriate number of target PW bands (energy range) to match, and thus the size of the Hilbert space, or to achieve convergence for systems with diffused electrons.

\begin{figure}[h]
\includegraphics[width=86mm]{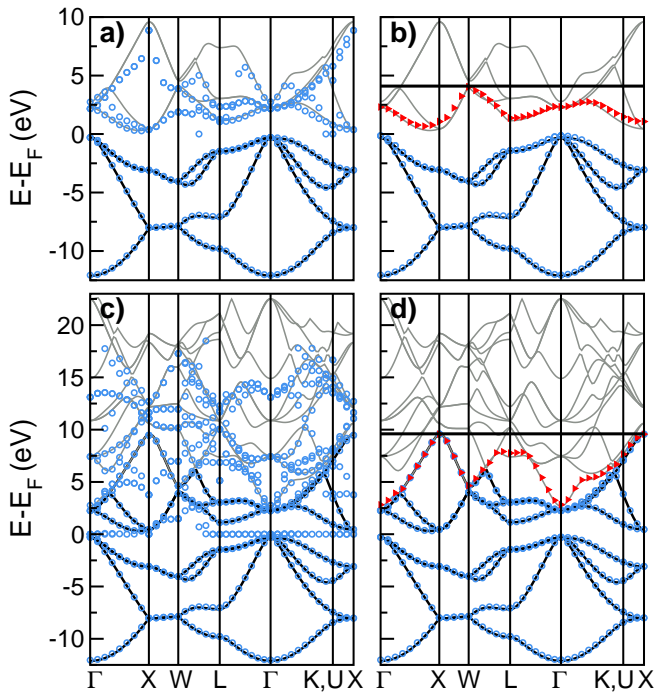}%
\caption{\label{fig:fig_si} (Color online) Si bandstructures from projected LCAO Hamiltonian matrices $H^{\mathbf{k}}(\kappa,N)$. Eigenenergies are marked by circles, triangles, or solid lines (flat bands). The panels on the top (bottom) row show the electronic structure expanded in a $sp$ ($spd$) finite space.
(a and c) \textit{Direct-projection} scheme (no filtering $N$=$M$, no shifting $\kappa$=0).
(b and d) Filtered+shifted projections. The degenerate flat bands (3- and 10-fold) are rigidly shifted by $\kappa$ (4.1 eV and 9.6 eV), and plotted as horizontal solid lines.
The reference PW bands are shown in black ($\mathcal{P}_n \geq 0.9$) or gray ($<0.9$).}
\end{figure}

On the other hand, noniterative methods such as \textit{direct projection} (Ref.~\onlinecite{Marzari2012}, II.i.1), \textsc{quambo} \cite{Lu2004}, and \textsc{qo} \cite{Qian2008}, which do not seek an iterative construction of the finite Hilbert space but rely on an AO basis provided as input, are a good compromise between speed (noniterative post-processing) and accurate reproduction of the occupied energy bands while keeping AO similarity. 

While the primary goal of these methods is to construct real-space wavefunctions with high localization, reciprocal-space Hamiltonians matrices can be built using these wavefunctions as well. The resulting band structure can exactly reproduce unentangled bands (bundles of bands that do not overlap others in energy across the BZ and are, therefore, isolated by energy gaps), e.g. the occupied manifold of an insulator or a semiconductor like silicon [black curves in Fig. 1(a)]. In general, however, the bands in metals (or the unoccupied manifold in semiconductors) are entangled. While the iterative MLWF method can further enforce reproduction of those bands inside a \textit{frozen} energy window via a disentanglement procedure \cite{Souza2001},  adding an extra layer of computational complexity, the noniterative methods fail to systematically reproduce those bundles, especially the upper unoccupied states [circles in Fig. 1(a)]. 

In this article, we present a fast, noniterative procedure that effectively and accurately reproduces a \textit{predictable} number of eigenenergies (bands) regardless of entanglement. The antibonding fraction of the wavefunction, scattered through the quasi-\textit{infinite} unoccupied subspace of the PW approach, is efficiently mapped into the finite-space Hamiltonian through straightforward matrix operations of projections on controlled basis and sequential filtering to automatically discard all unphysical solutions.

\section{methodology}
\subsection{LCAO representation of Bloch states}
For simplicity, in this work we use the finite Hilbert spaces defined by the set of pseudo atomic orbitals (PAOs) $\phi(\mathbf{r})$, employed in the generation of atomic pseudopotentials. 
While the richness of the PAO basis can be systematically increased by including more radial functions and angular momentum projectors in the construction of pseudo-wavefunctions, we stress that this is just a choice of convenience. Our procedure is completely general and can be applied to any finite basis, including polarized and diffused Gaussian sets, thus providing a direct bridge between the languages of solid-state physics and theoretical quantum chemistry.

As a first example we choose silicon, (see Fig. 1), where we have used a $sp$ (minimal) basis set
to construct the Bloch sums, 
$\phi_{\mu\mathbf{k}}(\mathbf{r}) = \frac{1}{\sqrt{N_{\mathcal{V}}}} \sum\limits_{\mathbf{R}} e^{i\mathbf{k}\cdot\mathbf{R}}\phi_\mu(\mathbf{r}-\mathbf{R})$, that span the finite Hilbert space $\Omega_{1}$
\footnote{\label{ftn:note1} The PAOs are from publicly available files. $\Omega_1$:$[\text{Ne}]\,3s^2 3p^2$, Si.pbe-n-rrkjus\_psl.0.1.UPF; \\$\Omega_2$:$[\text{Ne}]\,3s^2 3p^2 3d^0$, Si.pbe-n-rrkjus\_psl.0.1.UPF; \\ $\Omega_3$:$[\text{core}]\,4s^2 4p^6 5s^1 4d^5 5p^0$, Mo.pbe-spn-rrkjus\_psl.0.2; \\ $\Omega_4$:$[\text{core}]\,5d^{9.5} 6s^1 6p^{0.5}$, Au.pbe-van\_ak.UPF;\\ at \texttt{http://www.quantum-espresso.org/?page\_id=190}}.

These Bloch sums can be seen as the discrete Fourier transform of the corresponding PAO $\phi_{\mu}(\mathbf{r})$ replicated on a periodic box containing $N_{\mathcal{V}}$ lattice vectors $\mathbf{R}$ and, thus, the same number of $\mathbf{k}$-vectors in the BZ.
The starting PW Bloch states $|\psi_{n\mathbf{k}}^{PW}\rangle$ are obtained using the \textsc{quantum espresso} packages \cite{Giannozzi2009}. For convenience, let us switch to a L\"owdin-orthogonalized basis representation
$\bar{\phi}_{\mu\mathbf{k}}=\sum\limits_{\nu}({S^{\mathbf{k}}}^{-\frac{1}{2}})_{\mu\nu}\phi_{\nu\mathbf{k}}$ 
where 
$S^{\mathbf{k}}_{\mu \nu}=\langle \phi_{\mu\mathbf{k}} | \phi_{\nu\mathbf{k}}\rangle$ 
are the overlap matrices. The PW states are projected onto the finite Hilbert space of the PAOs via the operator 
$\hat{P}^{\mathbf{k}}= \sum\limits_{\mu \nu}|\bar\phi_{\mu\mathbf{k}}\rangle \langle \bar\phi_{\nu\mathbf{k}}|$.
Then, $|\psi_{n\mathbf{k}} \rangle = \sum\limits_{\mu} a_{\mu n}^{\mathbf{k}}|\bar{\phi}_{\mu\mathbf{k}}\rangle$
with expansion coefficients
\begin{equation*}
a_{\mu n}^{\mathbf{k}}=
\sum\limits_{\nu p}({S^{\mathbf{k}}}^{-\frac{1}{2}})_{\mu\nu}(S^{\mathbf{k}}_{\nu p})^{-1} \langle \phi_{p\mathbf{k}}|\psi_{n\mathbf{k}}^{PW}\rangle.
\end{equation*}
Similar expansions are used in the linear-combination-of-atomic-orbitals (LCAO) solution of the Kohn-Sham equation \cite{Dovesi2005}, where the Hamiltonian is 
\begin{equation} \label{eq:AEK}
\mathcal{H}^{\mathbf{k}}=A^{\mathbf{k}} {E}^{\mathbf{k}} A^{\mathbf{k}}{}^\dagger
\end{equation} 
and the matrices $A^{\mathbf{k}}$ of the expansion coefficients are found self consistently under the orthonormality constraint $A^{\mathbf{k}\dagger}A^{\mathbf{k}}=I$. 
Here $A^{\mathbf{k}}$ is built columnwise from the projection coefficients $a_{\mu \nu}^{\mathbf{k}}$; each column represents the LCAO wavefunction $|\psi_{n\mathbf{k}}\rangle$ for a given band $n$.
$\mathcal{H}^{\mathbf{k}},A^{\mathbf{k}}$ are of dimensions $M \times M, M \times N$, respectively. $M$ is the size of the Hilbert space and $N$ is the number of PW bands selected for projection. $E^{\mathbf{k}}$ is a $N \times N$ diagonal matrix of the $N$ lowest PW eigenvalues, $E^{\mathbf{k}}=\text{diag}(\epsilon^{PW}_1,\epsilon^{PW}_2,...,\epsilon^{PW}_N)$. 

\subsection{Band projectability}
Going back to the example of Si in space $\Omega_1$ ($M$=8) of Fig. 1(a), its corresponding LCAO Hamiltonian $\mathcal{H}^{\mathbf{k}}$ yields a manifold of 8 bands (marked with circles), some of which, especially in the unoccupied energy region, noticeably deviate from the reference PW bands (gray lines). 
This stems from the less than perfect projectability of those unoccupied PW states onto the inherently-incomplete finite space.
In turn, low projectability breaks the unitary constraint of the LCAO method, ${A}^{\mathbf{k}} {A} ^{\mathbf{k}}{}^\dagger\ne1$ . 
To quantify this effect, we define a projectability number 
$\mathcal{P}_{n}=\text{min}\{ \sum\limits_{\mu} a_{\mu n}^{\mathbf{k}*}a_{\mu n}^{\mathbf{k}},\,\forall \mathbf{k} \in\text{BZ} \}$
as an \textit{a priori} test for the representability of each PW band.
The closer $\mathcal{P}_n$ to 1, the better the fidelity of the corresponding LCAO band. Numerical values of  $\mathcal{P}_n$ for the systems studied here are summarized in Table~\ref{tab:table1}. We set an arbitrary cutoff of 0.9 as the condition for \textit{exact} representability so that with the minimal space $\Omega_1$ we can expect exact representability only for the lowest $N$=4 PW bands (the complete occupied manifold [black lines in Fig.~\ref{fig:fig_si}(a)]) with deviations less than 57.2 meV. A richer space $\Omega_{2}$ ($M$=18 \footnotemark[1]), which includes $d$ functions, yields $\mathcal{P}_7 \geq 0.9$ and should therefore support exact representation of the electronic structure up to $N$=7 PW bands [black lines in Fig.~\ref{fig:fig_si}(c)], which includes 3 unoccupied bands. However, the apparently bad reproducibility of the electronic structure in the energy window 0--10 eV observed in Fig.~\ref{fig:fig_si}(c) seems to suggest otherwise.

The progressive loss of representability of the top unoccupied PW bands mentioned above is a common issue even for more sophisticated noniterative and iterative approaches and not at all limited to the LCAO bands obtained as first-order solutions via the \textit{direct-projection} scheme. 
It is nonetheless observed that: 
(\textit{i}) The overall matching to the lower unoccupied bands significantly improves with the richer space $\Omega_2$ compared to $\Omega_1$ (cf. circles on solid lines for the 1st unoccupied band along the $\Gamma$-K path and along $\Gamma$-X for the 2nd and 3rd).
(\textit{ii}) Unoccupied states increasingly exhibit high-kinetic-energy plane-wave components, which do not project well on localized basis. The low projectability of the upper PW states yields low values of the $a_{mn}^{\mathbf{k}}$ coefficients. Consequently, the corresponding LCAO eigenstates consistently default to regions of lower energy and yield the apparent overall poor reproducibility seen in Fig.~\ref{fig:fig_si}(c).
(\textit{iii}) Singular values are introduced in the Hamiltonian in the case of no projectability, thus the observed zero-energy LCAO eigenvalues.

\subsection{Band filtering and shifting} \label{sec:filtering}
In order to resolve points (\textit{ii})  and (\textit{iii}) we filter out PW states with low projectability by choosing all the PW bands that satisfy $\mathcal{P}_n \geq 0.9$. The number of these PW bands, $N$, determines the number of columns of the matrix $A^{\mathbf{k}}$, which is  generally not square and non unitary.

In the formulation of the LCAO Hamiltonian we can always define a square $M \times M$ matrix $[A^{\mathbf{k}} \mathbf{0}]$, which extends $A^{\mathbf{k}}$ with $M-N$ columns of zeros, and fill ${E}^{\mathbf{k}}$ with the same number of zero eigenvalues. Then, Eq.~(\ref{eq:AEK}), can be more generally written as a product of square matrices 
\begin{equation} \label{eq:Hk2}
\mathcal{H}^{\mathbf{k}}=[A^{\mathbf{k}} \textbf{0}]\text{diag}(\epsilon^{PW}_1,...,\epsilon^{PW}_N,0,...,0)[A^{\mathbf{k}} \textbf{0}]^\dagger. 
\end{equation}

Since the product $[A^{\mathbf{k}} \textbf{0}][A^{\mathbf{k}} \textbf{0}]{}^{\dagger}$ is not unitary, the spectrum of its eigenvalues will contain 
the first N eigenvalues of $A^{\mathbf{k}} A^{\mathbf{k}}{}^\dagger$, of the form $\lambda_n=1-\delta_n$, with $0 \le \delta_n \le 0.1$ guaranteed by the projectability criterion adopted. The remaining $M-N$ eigenvalues will all be zero.

Using the canonical representation, $[A^{\mathbf{k}} \textbf{0}][A^{\mathbf{k}} \textbf{0}]^\dagger=UDU^\dagger$ we can write $[A^{\mathbf{k}} \textbf{0}]=UD^{\frac{1}{2}}$, where $D=\text{diag}(\lambda_1,...,\lambda_N,\lambda_{N+1},...,\lambda_M)$, and rewrite Eq.~(\ref{eq:Hk2}) as:
\begin{equation*}  \label{eq:Hk3}
\mathcal{H}^{\mathbf{k}}=U\text{diag}(\epsilon^{PW}_1(1-\delta_1),...,\epsilon^{PW}_N(1-\delta_N),0,...,0)U^\dagger.
\end{equation*}
This expression explicitly shows that having a non unitary space $A^{\mathbf{k}}$ applies a multiplicative factor to each PW eigenvalue. Filtering the low-projectability bands is a necessary step to guarantee that these corrections are minimal.
The $M-N$ null eigenvalues appear as degenerate ``flat'' bands at zero energy and need to be selectively moved out of the energy window of interest. 
To do so we introduce the ``shifting'' operator:
 $$\mathcal{I}^{\mathbf{k}} =I-A^{\mathbf{k}}A^{\mathbf{k}}{}^\dagger = U\text{diag}(\delta_1,...,\delta_N,1,...,1)U^\dagger$$
and rewrite the Hamiltonian as:
\begin{equation}\label{eq:central}
H^{\mathbf{k}}(\kappa,N)=\mathcal{H}^{\mathbf{k}}+\kappa\mathcal{I}^{\mathbf{k}}= U\text{diag}(\bar{\epsilon}_1,...,\bar{\epsilon}_N,\kappa,...,\kappa)U^\dagger.
\end{equation}

Eq.~(\ref{eq:central}) is the central result of our work. The $\mathcal{I}^{\mathbf{k}}$ operator allows a selective shifting, by $\kappa$, of the $M-N$ flat bands with negligible effect on the other $N$ high-projectability bands, with eigenenergy deviations  $\Delta \epsilon_n \le (1-\mathcal{P}_n)(\kappa - \epsilon^{PW}_n)$. The last factor is interpreted as the energy window between the highest band under consideration ($N$th) of energy $\sim\kappa$ and a given PW eigenvalue, and is maximal for the lowest band $n$=1. In practice, since the upper limit of the energy window is usually only $\sim$2-5 eV above E${}_{\text{F}}$, the correction can be made arbitrarily small by an appropriate choice of the projectability number $\mathcal{P}_n$.

\begin{table}[b]
\caption{\label{tab:table1}
Projectability ($\mathcal{P}_n$) and BZ rms average of $\Delta\epsilon_n$ (in eV) per band $n$ of $H^{\mathbf{k}}(\kappa,N)$. $\kappa=0,9.6,10$ eV; $N=5,8,13$ for Si - $\Omega_{1}$, Si - $\Omega_{2}$, and Mo - $\Omega_{3}$, respectively} 
\begin{ruledtabular}
\begin{tabular}{ccccc|ccc}
& \multicolumn{2}{c}{Si - $\Omega_{1}$} 
& \multicolumn{2}{c}{Si - $\Omega_{2}$} 
& \multicolumn{3}{c}{Mo - $\Omega_{3}$\footnotemark[1]} \\ 
$n$ & $\mathcal{P}_n$ & rms & $\mathcal{P}_n$ & rms & $n$ & $\mathcal{P}_n$ & rms \\ \hline
1 & 0.9927& 0.0478&  0.9953&0.0792& 5 &0.9979 & 0.0140 \\ 
2 & 0.9622& 0.0572&  0.9933&0.0555& 6 &0.9937 & 0.0183 \\
3 & 0.9622& 0.0252&  0.9952&0.0382& 7 &0.9873 & 0.0386 \\
4 & 0.9622& 0.0203&  0.9967&0.0266& 8 &0.9924 & 0.0275 \\
5 & 0.4755& 1.9200&  0.9934&0.0267& 9 &0.9905 & 0.0302 \\
6 & 0.0660&       &  0.9771&0.0316&10 &0.9286 & 0.0602 \\
7 & 0.0941&       &  0.9728&0.0380&11 &0.0000 & 3.5975 \\
8 & 0.0652&       &  0.8464&0.1382&12 &0.0000 & 3.3524 \\
\end{tabular}
\end{ruledtabular}
\footnotetext[1]{The 4 semicore bands have exact projectability.}
\end{table}
The LCAO  bandstructures (circles and triangles) in Fig. 1(b) and (d) showcase the dramatic improvement, with respect to (a) and (c), that is achieved by filtering and shifting the low-projectability bands. It confirms the exact reproductibility of 4 and 7 bands (circles) expected for the two different spaces, respectively. For further illustration, we included the projections of an additional PW band (with $\mathcal{P}_n< 0.9$) in building the Hamiltonians.
As expected, this LCAO band [5th in (b) and 8th in (d), marked with triangles] does not \textit{exactly} reproduce the PW reference. Nonetheless, the band closely follow the reference across the BZ, except at X and U-X in (b) and L in (d), which are the BZ regions where the projectability is the lowest. The 8th band shows an overall good fidelity, consistent with the relatively high projectability $\mathcal{P}_8 = 0.8335$.

\begin{figure}[ht]
\includegraphics[width=86mm]{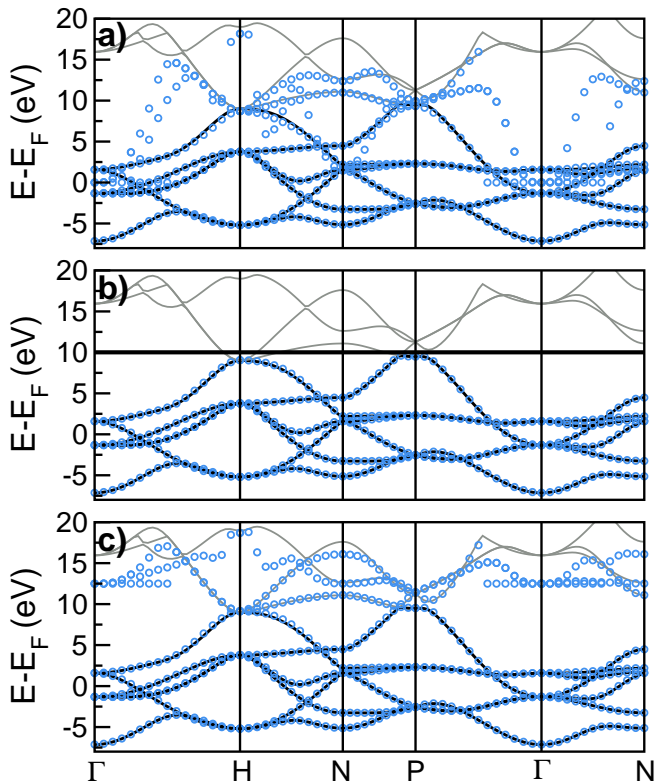}%
\caption{\label{fig:fig_bccmo} (Color online) Bandstructure of Mo bcc under space $\Omega_3$. (a) \textit{Direct-projection} scheme ($N$=13, $\kappa$=0 eV). (b) Filtered+shifted projection ($N$=7, $\kappa$=10 eV) (c) Unfiltered+shifted projection ($N$=$M$=13, $\kappa$=12.5 eV). The eigenenergies of the LCAO Hamiltonian $H^{\mathbf{k}}(\kappa,N)$ are shown with blue circles. The PW bands of high $\mathcal{P}_n\ge0.9$ (low $\mathcal{P}_n<0.9$) projectability are shown with solid black (gray) lines. The 4 low-lying semicore bands ($4s4p$) are not shown. Inset shows the Fermi surface in its Wigner-Seitz cell.}
\end{figure}

To demonstrate the performance of our procedure in the case of entangled bands, we computed the band structure of  an intrinsically delocalized metallic systems, such as molybdenum bcc (Fig. 2, $\Omega_3$ \textit{M}=13 \footnotemark[1]) and of a gold nanowire (Fig.~\ref{fig:fig_au_nano}, $\Omega_4$ \textit{M}=90 \footnotemark[1]).
As observed in the silicon case, in both metallic systems the low-projectability states can default to the bottom of the unoccupied energy region, as in the \textit{direct-projection} bandstructure in Fig. 2(a) for Mo. The effect of the low-projectability states is more detrimental in the nanowire case where they hybridize with states of otherwise high fidelity and any resemblance to the PW reference is lost [Fig.~\ref{fig:fig_au_nano}(a)].

For Mo, the space $\Omega_3$ supports exact representation of up to 6 PW bands (beyond the 4 semicore bands, not shown), as found in Table~\ref{tab:table1} for $\mathcal{P}_n\geq0.9$. The bandstructure from the filtered+shifted scheme [Fig.~\ref{fig:fig_bccmo}(b)] confirms reproducibility of all these PW bands with root-mean-square (rms) deviations less than 60.2 meV even for the large value of $\kappa$ (10 eV) used here. 

In particular cases it may be advantageous to keep the low-projectability PW states as in the \textit{direct-projection} scheme; for instance, the low-projectability PW bands $n$=11 and 12 of Mo bcc exhibit \textit{local} high projectability around H and P. Applying a $\kappa$ shift to the \textit{direct-projection} Hamiltonian, i.e. $H^{\mathbf{k}}(12.5,M)$, 
yields an expanded range of bandstructure reproducibility of up to 12.5 eV [Fig. 2(c)], which is directly due to the inclusion of bands 11 and 12.

\section{Applications of LCAO Hamiltonians}

\subsection{Band interpolation}
Once the LCAO matrices $H^{\mathbf{k}}(\kappa,N)$ are known, one can directly construct the real-space localized Hamiltonian as 
\begin{equation} \label{eq:ft}
H^{\mathbf{0R}}=\frac{1}{\sqrt{N_{\mathcal{V}}}} \sum\limits_{\mathbf{k}} e^{-i\mathbf{k}\cdot\mathbf{R}}H^{\mathbf{k}}(\kappa,N).
\end{equation}
Conversely, these local matrices allow us to obtain the \textit{interpolated} bandstructure at any arbitrary $\mathbf{k}$-vector ~\cite{Agapito2007, Lee2005}, with the same accuracy defined by the projectability number, by diagonalizing the \textit{interpolated} reciprocal-space Hamiltonian 
\begin{equation*}
H^{\mathbf{k},\textit{interpolated}}= \sum_{\mathbf{R}} e^{i\mathbf{k}\cdot\mathbf{R}}H^{\mathbf{0R}}.
\end{equation*}

\subsection{Band decomposition and Fermi surface analysis}
\begin{figure}[ht]
\includegraphics[width=86mm]{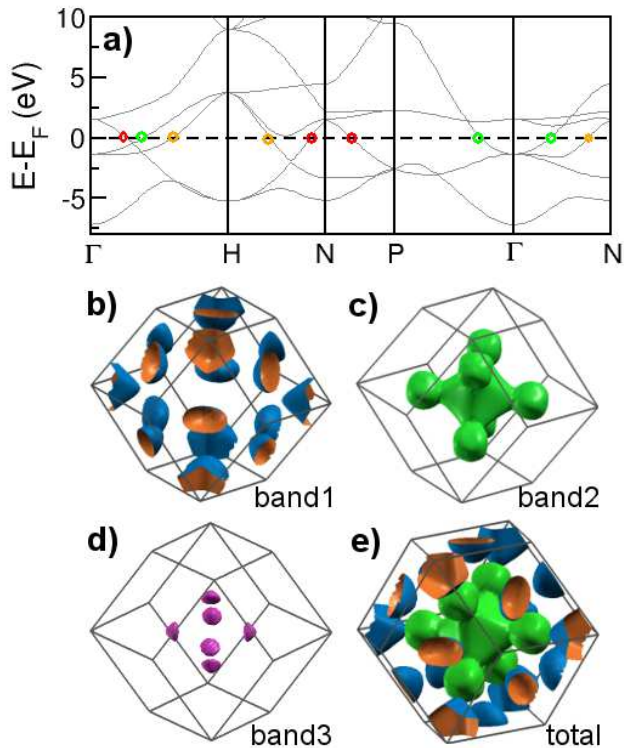}%
\caption{\label{fig:fig_fs} (Color online) Fermi surface of Mo bcc bulk. (a) Colored circles identify the same band crossing the Fermi energy (dashed horizontal line). Orange=band 1, green=band 2, red=band 3. (b-d) Individual-band contributions to the Fermi surface. (e) Total Fermi surface.} 
\end{figure}

A valuable application of this procedure is in the evaluation of fundamental physical properties of materials that require an accurate representation of the electronic states across the whole BZ. A typical example is the calculation of the Fermi surface of any metal, which typically requires extraordinary computational efforts. Within our approach the Fermi surface is straightforwardly obtained by the direct evaluation of the \textit{interpolated} bandstructure of the system via the real-space Hamiltonians. 
The Fermi energy of Mo bcc is crossed at various $\mathbf{k}$-points in the 3-dimensional BZ. The collection of all such points define its Fermi surface, which is shown in Fig.~\ref{fig:fig_fs}(e). 
Furthermore, it is determined that the crossing states can have three distinct atomic characters. 
Bands of different character are marked with circles of different colors at the crossing of the Fermi level (horizontal dashed line) in Fig.~\ref{fig:fig_fs}(a).
The decomposition of the Fermi surface based on the atomic character of the crossing states is shown in Fig.~\ref{fig:fig_fs}(b)-(d).

\subsection{Electrical conductance}
\begin{figure}[h]
\includegraphics[width=86mm]{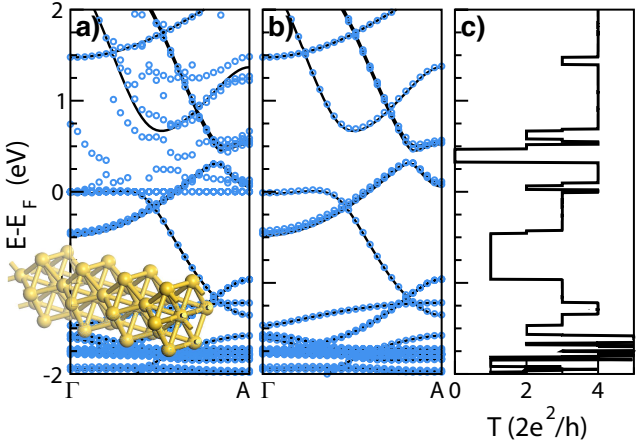}%
\caption{\label{fig:fig_au_nano} (Color online) Bandstructure and quantum transmittance spectrum of a gold nanowire. 
(a) The \textit{direct-projection} and (b) filtered+shifted ($N$=62, $\kappa$=3 eV) bandstructures are marked with circles. Solid lines are PW bands.
(c) Quantized transmittance of the nanowire.}     
\end{figure}

Another straightforward application of LCAO Hamiltonians is the calculation of the electrical conductance through a nanowire. Our procedure reduces the problem of calculating electron transport \cite{Agapito2007, Calzolari2004} to the computationally inexpensive post-process of evaluating Eq.~(\ref{eq:tf}). 
We choose a gold nanowire \cite{Krstic2002} as a prototypical example.
Bands obtained from the \textit{direct-projection} scheme are shown in Fig.~\ref{fig:fig_au_nano}(a). Unphysical zero-energy eigenstates, discussed in Section~\ref{sec:filtering}, cover the entirety of the BZ. They are removed by applying a shifting of $\kappa = 3$ eV to the filtered bands (lowest $N$=62 bands) and the resulting LCAO eigenstates are reported in Fig.~\ref{fig:fig_au_nano}(b) with blue circles. The LCAO real-space Hamiltonians $H^{\mathbf{00}}$ and $H^{\mathbf{01}}$ are obtained according to Eq.~(\ref{eq:ft}) and the electrical transmittance $T(E)$, in Fig.~\ref{fig:fig_au_nano}(c), is computed from these matrices as described in the Appendix.

\begin{figure}[h]
\includegraphics[width=86mm]{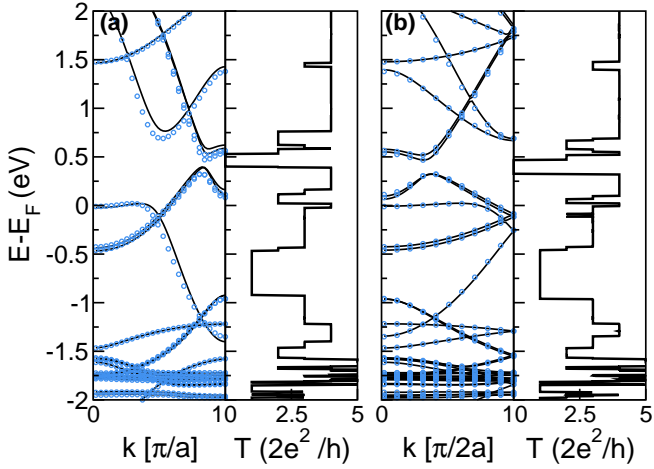}%
\caption{\label{fig:fig_au_nano2} (Color online) Effect of the nearest principal-layer approximation on the interpolated bandstructure (left) and quantized conductance (right). The principal layer is approximated with one and two unit cells (\textit{a}=4.71 \AA) in panels (a) and (b), respectively. The blue circles are the eigenvalues of $H^{\mathbf{k}}$ and the underlying solid lines are the corresponding \textit{interpolated} bands.}
\end{figure}

Furthermore, formulating the LCAO Hamiltonian $H^{\mathbf{k}}$ directly in reciprocal space allows the discrimination of the two most common sources of error in electronic transport simulations: (\textit{i}) the incompleteness of the finite space, which is reflected in the mismatch between the PW and LCAO bandstructures [cf. Fig.~\ref{fig:fig_au_nano2}(b)] and can be made arbitrarily small by our procedure; and (\textit{ii}) the real-space interaction truncation in the choice of the principal layer. The latter is outside of the scope of any reciprocal-space mapping procedure $\psi_{n\mathbf{k}}^{PW} \rightarrow \psi_{n\mathbf{k}}^{LCAO}$ . Nonetheless, such error is introduced when formulating real-space local Hamiltonians $H^{\mathbf{0R}}$  [via Eq.~(\ref{eq:ft})]. This real-space truncation error is reflected in the mismatch between the \textit{interpolated} bandstructure (solid lines) and the LCAO eigenvalues (blue circles) seen in Fig.~\ref{fig:fig_au_nano2}(a) and can be systematically reduced by increasing the size of the principal layer. For instance, the mismatch is eliminated when doubling the size of the nanowire principal layer (to a lattice constant of $2\times4.71$ \AA), as shown in Fig.~\ref{fig:fig_au_nano2}(b).

The three applications shown here are implemented in the GPL open-software packages \textsc{WanT} and \textsc{quantum espresso} of Refs. 4 and 19, respectively.

\section{conclusions}
These results have far reaching implications, well beyond the practical applications shown here. In fact, our method allows control of the size (richness) of the finite Hilbert space of the basis functions, an archetypal feature in quantum chemistry, which in turns provides flexibility to converge to the \textit{true} \textit{infinite}-space solution in the limit of infinite plane waves \footnote{\label{ftn:note3}L. Agapito and M. Buongiorno Nardelli, \textit {in preparation}}. 
Expanding the space increases the number of unoccupied bands with the concomitant bad reproducibility problem that challenges current methodologies and has biased solutions toward minimal basis, which only bypasses the problem by reducing the number of unoccupied bands. 

Contrarily to the spirit of the \textit{N}MTO and MLWF methods, our technique does not seek construction of (heavily customized, localized) basis functions. Its importance resides on allowing noniterative reproduction a large number of energy bands using standard quantum-chemistry basis sets or equivalent. In that regards, the present methodology completely supersedes the need for engineered basis functions such as MLWFs or \textit{N}MTOs.

The central result of this article, Eq.~(\ref{eq:central}), allows us to obtain an exact representation of all the PW bands that complies with our high projectability criterion in a procedure of negligible computational cost, opening the way to the design of efficient algorithms for electronic structure simulations of realistic material systems, within the high-throughput materials framework \cite{aflowlibPAPER}.

\begin{acknowledgments}
This work was supported, in part, by SRC through Task ID P14924 in the Center for Electronic
Materials Processing and Integration at the University of North
Texas, and by the Duke University, Center for Materials Genomics.
\end{acknowledgments}

\appendix*
\section{Electrical conductance in periodic nanowires}
The set of real-space Hamiltonian $H^{\mathbf{0R}}$ and overlap matrices $S^{\mathbf{0R}}$, in the general case of nonorthogonal basis, contain the necessary information to compute the electronic properties along the nanowire.

In the site representation, the spatial coordinate $\mathbf{r}$ of the wavefunction is discretized to lattice vector $\mathbf{R}$, thus, each periodic cell is considered an abstract single site $\mathbf{R}$. 

Following the translational Bloch theorem, a wavefunction $\psi^{\mathbf{k}} \equiv \psi^{\mathbf{k}}_{\mathbf{0}}$ evaluated at a generic site $\mathbf{r}=\mathbf{0}$ relates to the next site $\mathbf{r}=\mathbf{-1}$ or $\mathbf{1}$ by a phase factor, that is
$\psi^{\mathbf{k}}_{\mathbf{-1}}=e^{-i\theta} \psi^{\mathbf{k}}_{\mathbf{0}}$ and 
$\psi^{\mathbf{k}}_{\mathbf{1}}=e^{i\theta} \psi^{\mathbf{k}}_{\mathbf{0}}$, 
where $\theta = \mathbf{k}\cdot\mathbf{1}$.

In a nonorthogonal LCAO space, the wavefunction satisfies the Roothaan equation
\begin{equation} \label{eq:schrod}
H^{\mathbf{k}}\psi^{\mathbf{k}}=E^{\pm} S^{\mathbf{k}}\psi^{\mathbf{k}}
\end{equation}

Both Hamiltonian and overlap matrices are obtained from the Fourier transform or the corresponding real-spaces quantities.
\begin{equation*}
H^{\mathbf{k}}= \sum_{\mathbf{R=-1,0,1}} e^{i\mathbf{k}\cdot\mathbf{R}}H^{\mathbf{0R}}
\end{equation*}
\begin{equation*}
S^{\mathbf{k}}= \sum_{\mathbf{R=-1,0,1}} e^{i\mathbf{k}\cdot\mathbf{R}}S^{\mathbf{0R}}  
\end{equation*}
The truncation in the Fourier transform corresponds to the principal-layer (PL) approximation. A PL is composed of one or more primitive cells, such that interactions beyond nearest PLs are made negligible. Then, Eq.~(\ref{eq:schrod}) becomes 
\begin{equation} \label{eq:quadratic1}
\left[ M\lambda -h - h^{t}\lambda^2 \right] \psi^{\mathbf{k}}_{\mathbf{1}} =0
\end{equation}
or equivalently,
\begin{equation} \label{eq:quadratic2}
\left[ M\lambda^{-1} -h^{t} - h\lambda^{-2} \right] \psi^{\mathbf{k}}_{\mathbf{-1}} =0 
\end{equation}

with $\lambda = e^{-i\theta}$; $M^{\pm}=E^{\pm}S^{\mathbf{00}}-H^{\mathbf{00}}$; $h^{\pm}=H^{\mathbf{01}}-E^{\pm}S^{\mathbf{01}}$, and the definition $h^{t}(E) \equiv h^{\dagger}(E^{*})$. It assumes hermitian matrices, i.e. $H^{\mathbf{0,-1}}=(H^{\mathbf{01}})^{\dagger}$.

Eqs.~(\ref{eq:quadratic1}) and (\ref{eq:quadratic2}) are standard quadratic eigenvalue problems of the form $a_2 \lambda^2 + a_1 \lambda + a_0 = 0$.
The solutions $\lambda_n$  represent all the propagating modes of the nanowire. Solutions with $|\lambda_n|>1$ are evanescent modes decaying (traveling) to the right while $|\lambda_n|<1$ are evanescent modes traveling to the left. Modes with $|\lambda_n|=1$ are standing waves, i.e. Bloch states. An infinitesimal imaginary quantity $\eta$ is added to the energy, thus the definition $E^{\pm}=E \pm i\eta$. In this way the phase factors are moved slightly away from the unitary circle and an unambiguous traveling direction can be assigned.

As a result, half of the 2$M$ solutions of the quadratic equation, are discerned as left traveling and the other as right traveling, denoted by subscripts $<$ and $>$, respectively.

Compounding all the ``left traveling'' eigenvalues $\lambda_{n<}$ and eigenvectors $\psi^{\mathbf{k}}_{\mathbf{1}n<}$ into the $M \times M$ matrices $\Lambda_{<}$ and $U_{1,<}$, respectively, the matrix $\alpha_{<}$ is defined as
\begin{equation} \label{eq:alpha}
\alpha_{<}=U_{1,<} \Lambda_{<} \left(U_{1,<} \right)^{-1},
\end{equation}

which satisfies Eq.~(\ref{eq:quadratic1}), i.e. $M\alpha-h-h^{t}\alpha^{2}=0$.

\tikzstyle{vecArrow} = [thick, decoration={markings,mark=at position
   1 with {\arrow[semithick]{open triangle 60}}},
   double distance=2.4pt, shorten >= 5.5pt,
   preaction = {decorate},
   postaction = {draw,line width=1.4pt, white,shorten >= 4.5pt}]
\begin{figure}[h]
\begin{center}
\begin{tikzpicture}[auto,node distance=10mm,
  thick,main node/.style={circle,fill=blue!15,minimum size = 5mm,inner sep = 0pt,draw},
  node2/.style={circle,fill=white,minimum size = 5mm,inner sep = 0pt}]
  \draw[dotted] (2,0) -- (4,0);
  \draw (-1,0) -- (2,0);
  \draw (4,0) -- (7,0);
  \node[main node] (1) {-3};
  \node[main node] (2) [right of=1] {-2};
  \node[main node] (3) [right of=2] {-1};
  \node[main node] (4) [right of=3] {0};
  \node[main node] (5) [right of=4] {1};
  \node[main node] (6) [right of=5] {2};
  \node[main node] (7) [right of=6] {3};
  \node[node2] (8) [left of=1] {};
  \node[node2] (9) [right of=7] {};
   \draw [bend left=70,->] (5) to node[midway] {$h$} (6);
   \draw [bend left=70,->] (6) to node {$h$} (7);
   \draw [bend left=70,->] (7) to node {$h$} (9);
   \draw [bend left=70,->] (7) to node {$h^{t}$} (6);
   \draw [bend right=70,->] (2) to node[above] {$h^{t}$} (1);
   \draw [bend right=70,->] (3) to node[above] {$h^{t}$} (2);
   \draw [bend right=70,->] (1) to node[above] {$h^{t}$} (8);
   \draw [bend right=70,->] (1) to node[below] {$h$} (2);
   \draw[vecArrow] (2.3,-0.5) to (2.9,-0.5);
   \draw[vecArrow] (3.7,-0.5) to (3.1,-0.5);
   \node at (2.6,-0.8) {$\beta_{>}$};
   \node at (3.5,-0.8) {$\alpha_{<}$};
  \path[every node/.style={font=\sffamily\small}]
   (4)   edge [loop above] node {$M$} (4);
\end{tikzpicture}
\end{center}
\caption{\label{fig:tb} (Color online) Schematic representation of the quadratic eigenvalue equations (\ref{eq:quadratic1}) and (\ref{eq:quadratic2}) as a 1-dimensional tight-binding model. }
\end{figure}
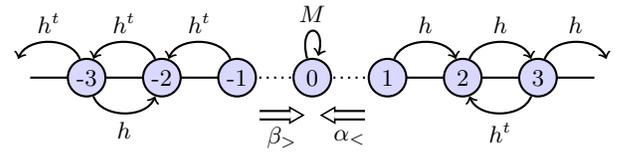

Physically, it represents the propagating modes of a semi-infinite right wire (starting at site $\mathbf{1}$) leaking into site $\mathbf{0}$, as schematically shown in Fig.~\ref{fig:tb}. Analogously, the effect of all right-moving modes of a semi-infinite left wire (which starts at site $\mathbf{-1}$) on site $\mathbf{0}$ is given by the solutions of Eq.~\ref{eq:quadratic2}. The matrix $\beta_{>}$ that satisfies that equation is 
\begin{equation} \label{eq:beta}
\beta_{>}=U_{-1,>} \left(\Lambda_{>}\right)^{-1}\left(U_{-1,>} \right)^{-1}
\end{equation}

The use of $h$ and $h^{t}$ is inverted in Eq.~(\ref{eq:quadratic2}) with respect to Eq.~(\ref{eq:quadratic1}). This effectively flips the wavevector of the modes, i.e. $\lambda \leftarrow \lambda^{-1}=\lambda^{*}$, and is accounted by taking the inverse of $\Lambda$ in Eq.~(\ref{eq:beta}).

Finally, in the case of an infinite nanowire, all relevant electronic properties such as the Green's function $G$, and the electrical conductance $T$ is directly obtained from $\alpha_{<}$ and $\beta_{>}$ using the following

\begin{equation*}
\Sigma_L^{\pm}=(h^\mp)^{\dagger} \alpha_{<}^{\pm}; \qquad
\Sigma_R^{\pm}=(h^\pm)\beta_{>}^{\pm}
\end{equation*}
\begin{equation*}
\Gamma_{L/R}=i\left(\Sigma_{L/R}^{+}-\Sigma_{L/R}^{-}\right)
\end{equation*}
\begin{equation*}
G^{\pm}=\left( h^{\pm} \left[ \left(\alpha_{<}^{\pm}\right)^{-1}-\beta_{>}^{\pm} \right] \right)^{-1}
\end{equation*}
\begin{equation} \label{eq:tf}
T(E)=trace\left[ \Gamma_L G^{+} \Gamma_R G^{-} \right]
\end{equation}

\bibliography{library}

\end{document}